\documentclass[aps,prl,showpacs,amsmath,amssymb,twocolumn,superscriptaddress]{revtex4-1}
\pdfoutput=1

\usepackage{graphicx}
\usepackage{color}
\usepackage{amsmath}

\renewcommand{\d}{\mathrm{d}}

\newcommand{\mr}[1]{\mathrm{#1}}
\newcommand{\refcite}[1]{Ref.~\onlinecite{#1}}

\begin{document}

\title{Manifestly non-Gaussian fluctuations in superconductor-normal metal tunnel nanostructures}

\author{M.~A.~Laakso}
\email[]{matti.laakso@aalto.fi}
\author{T.~T.~Heikkil\"a}
\affiliation{Low Temperature Laboratory, Aalto University, Post Office Box 15100, FI-00076 AALTO, Finland}
\author{Yuli V.~Nazarov}
\affiliation{Kavli Institute of Nanoscience, Delft University of Technology, 2628 CJ Delft, The Netherlands}
\date{\today}

\begin{abstract}
We propose a mesoscopic setup which exhibits strong and manifestly non-Gaussian fluctuations of energy and temperature when suitably driven out of equilibrium. The setup consists of a normal metal island (N) coupled by tunnel junctions (I) to two superconducting leads (S), forming a SINIS structure, and is biased near the threshold voltage for quasiparticle tunneling, $eV\approx2\Delta$. The fluctuations can be measured by monitoring the time-dependent electric current through the system, which makes the setup suitable for the realization of feedback schemes which allow to stabilize the temperature to the desired value.
\end{abstract}


\maketitle

In a (grand) canonical ensemble at temperature $T$, the total internal energy of the system fluctuates owing to the fluctuating energy and particle flows between the system and the thermal bath. It is a fundamental result of equilibrium statistical mechanics that the variance of energy is given by $\mr{Var}(E)=k_BC(T)T^2$, $C(T)$ being the (temperature dependent) heat capacity of the system \cite{reichl98}. For a system with a large number of degrees of freedom, the fluctuations are small and can be regarded as Gaussian. 

In many systems the internal relaxation rate is much faster than the rate of energy exchange with the environment. In this quasiequilibrium case, the probability distribution of the system is thermal with some effective temperature $T$, which is unambiguously related to the instantaneous total energy of the system $E$ via $\d E/\d T=C(T)$. The fluctuations of energy and effective temperature are thus the same. In a driven system, the energy is not necessarily Boltzmann-distributed, so that the corresponding ensemble is non-Gibbsian. Moreover, the fluctuations of energy/temperature are not generally Gaussian. However, in systems with a large number of degrees of freedom these properties are usually non-accessible: The effective temperature is close to its average value determined from the heat balance, and its fluctuations are small and Gaussian. Recently, fluctuation statistics of effective temperature have been studied in non-interacting electron islands \cite{heikkila09}, and overheated single-electron transistors \cite{laakso10,laakso10b}. Typically, the non-Gaussian effects are noticeable only for large and therefore exponentially improbable deviations from average values.

\begin{figure}
	\includegraphics[width=0.9\columnwidth]{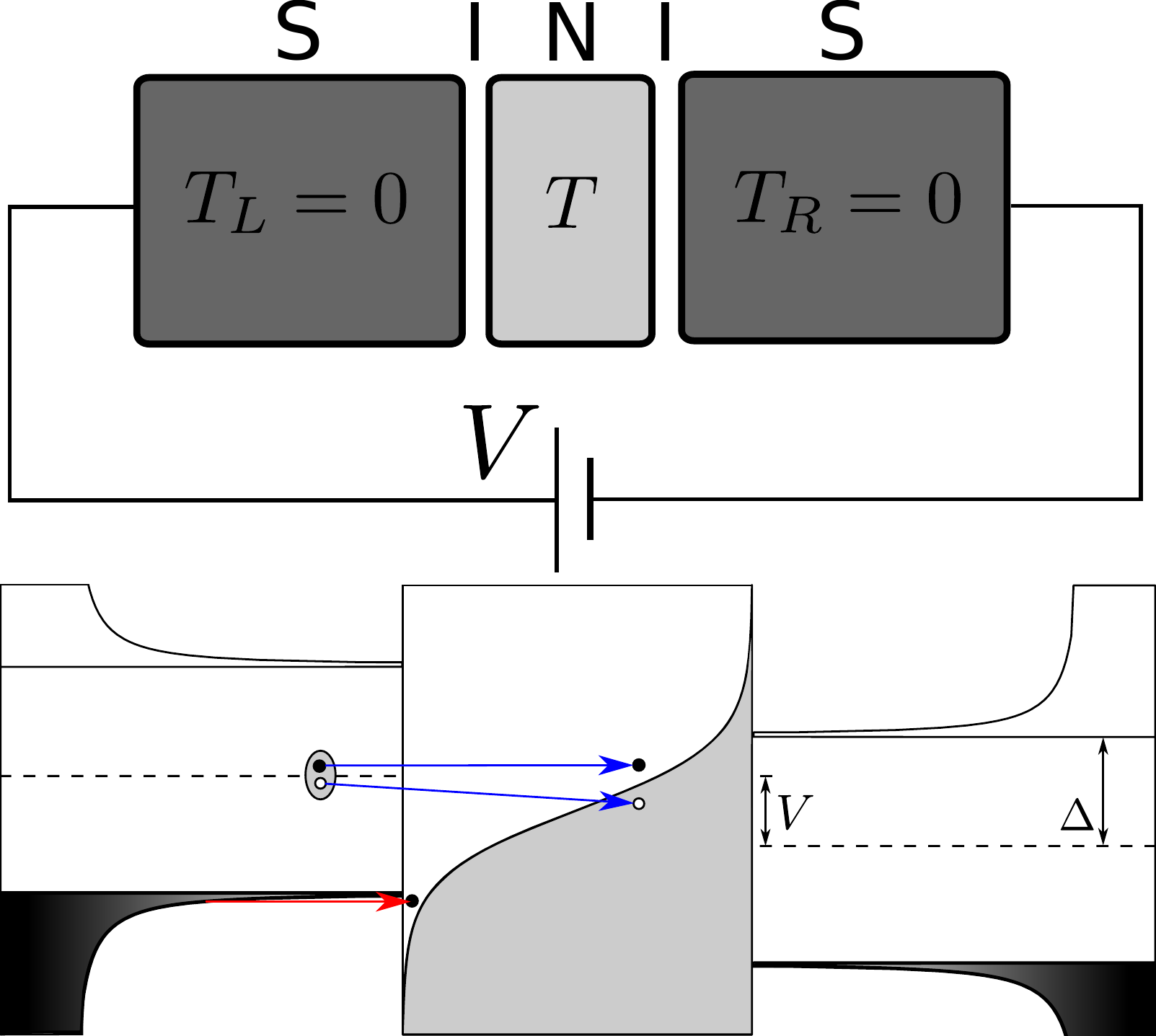}
	\caption{(color online) (top) SINIS structure biased at voltage $eV\approx2\Delta$. The normal metallic island is connected via tunnel contacts to superconducting reservoirs. The effective temperature on the island, $T$, fluctuates due to fluctuations in the energy flows to the leads. (bottom) Energy diagram of the system for $eV<2\Delta$, showing the BCS density of states in the leads and the Fermi distribution on the island. The first order processes (single red arrow) are thermally activated. The second order processes (two blue arrows) occur at all temperatures and correspond to incoherent tunneling of Cooper pairs.}
	\label{fig:schema}
\end{figure}

In this Letter we demonstrate the feasibility of strong fluctuations of temperature and manifestly non-Gaussian distribution of these fluctuations in a mesoscopic system with a large number of degrees of freedom. The system is a SINIS structure, shown schematically in Fig.~\ref{fig:schema}, where a normal metallic island is connected to two superconducting leads via tunnel junctions, and biased close to the threshold for quasiparticle tunneling, $eV\approx2\Delta$, $\Delta$ being the energy gap in the superconductors. The cause of these fluctuations is the interplay of regular quasiparticle tunneling and two-electron Andreev tunneling.

In this setup the temperature fluctuations can be easily and quickly monitored by measuring the electric current --- no separate thermometers are necessary. This permits a practical realization of a feedback scheme where the fluctuations are coupled to certain control parameters of the system, so-called Maxwell demons \cite{serreli07,horowitz11,pekola07}. For example, when the system fluctuates to a low temperature, the tunnel junctions could be "switched off," trapping the system at this temperature \footnote{For the duration of the electron--phonon scattering time, assumed long in this paper.}. In one parameter regime discussed below, it is possible to reach an extremely low effective temperature this way.

The basic physical mechanism responsible for the strong non-Gaussian fluctuations predicted here is the competition of quasiparticle and Andreev tunneling. Below and up to $eV=2\Delta$ the quasiparticle tunneling processes cool the island, each tunneling event extracting on average an energy of $k_BT \ll \Delta$. This is the well known cooling mechanism in SINIS structures \cite{nahum94,leivo96,giazotto06}. Andreev tunneling, a process where a Cooper pair in the superconductor is converted into two quasiparticles in the normal metal \cite{hekking93,rajauria08}, deposits a relatively large amount of energy, $2\Delta$, on the island. Energy and effective temperature are related by $E=\pi^2k_B^2T^2/(6\delta)$, $\delta\ll\Delta$ being the single-particle level spacing on the island, inversely proportional to its volume. A single Andreev event therefore heats the island to a temperature of at least $k_BT_t=\sqrt{12\delta\Delta}/\pi$. Owing to the heat balance, the rate of Andreev processes $\Gamma_A$ should be by a factor of $k_BT/\Delta$ smaller than the rate of quasiparticle tunneling. To this end, we may disregard the randomness in the quasiparticle flow and characterize the {\it deterministic} energy relaxation due to quasiparticles by a typical time $\tau_r$.



The regime of manifestly non-Gaussian fluctuations requires a low Andreev rate, $\Gamma_A \tau_r \lesssim 1$, implying \cite{belzig00} that all transmission eigenvalues $T_p$ of the junctions should be small, $\mathcal{F}\equiv\sum_p T^2_p/\sum_p T_p\ll1$ being a crucial parameter. This requirement is experimentally feasible: The Andreev events can be resolved in time \cite{maisi11}, and $\mathcal{F}\simeq 10^{-5}-10^{-6}$ for aluminum junctions \cite{greibe11}. Moreover, the reservoirs should be kept at a low temperature and the island should be small such that it can be cooled down to temperatures of the order of $T_t$, bringing the average {\it total} energy of the island down to the order of $\Delta$ \footnote{This is not a very limiting requirement, since the backflow of heat from the reservoirs at temperature $T_S$ is exponentially small for $k_BT_S\ll\Delta$, $\propto\sqrt{T_S}\exp[-\Delta/(k_BT_S)]$ \cite{anghel01}.}. In addition, we need to avoid the Coulomb blockade regime, so that the dimensionless conductance of the junctions should satisfy $g\equiv G/G_Q\gtrsim1$, $G_Q=e^2/(\pi\hbar)$. Finally, the heat exchange processes not involving electron transfers should be small enough not to disturb the competition between Andreev and quasiparticle events.

\begin{figure}
	\includegraphics[width=0.9\columnwidth]{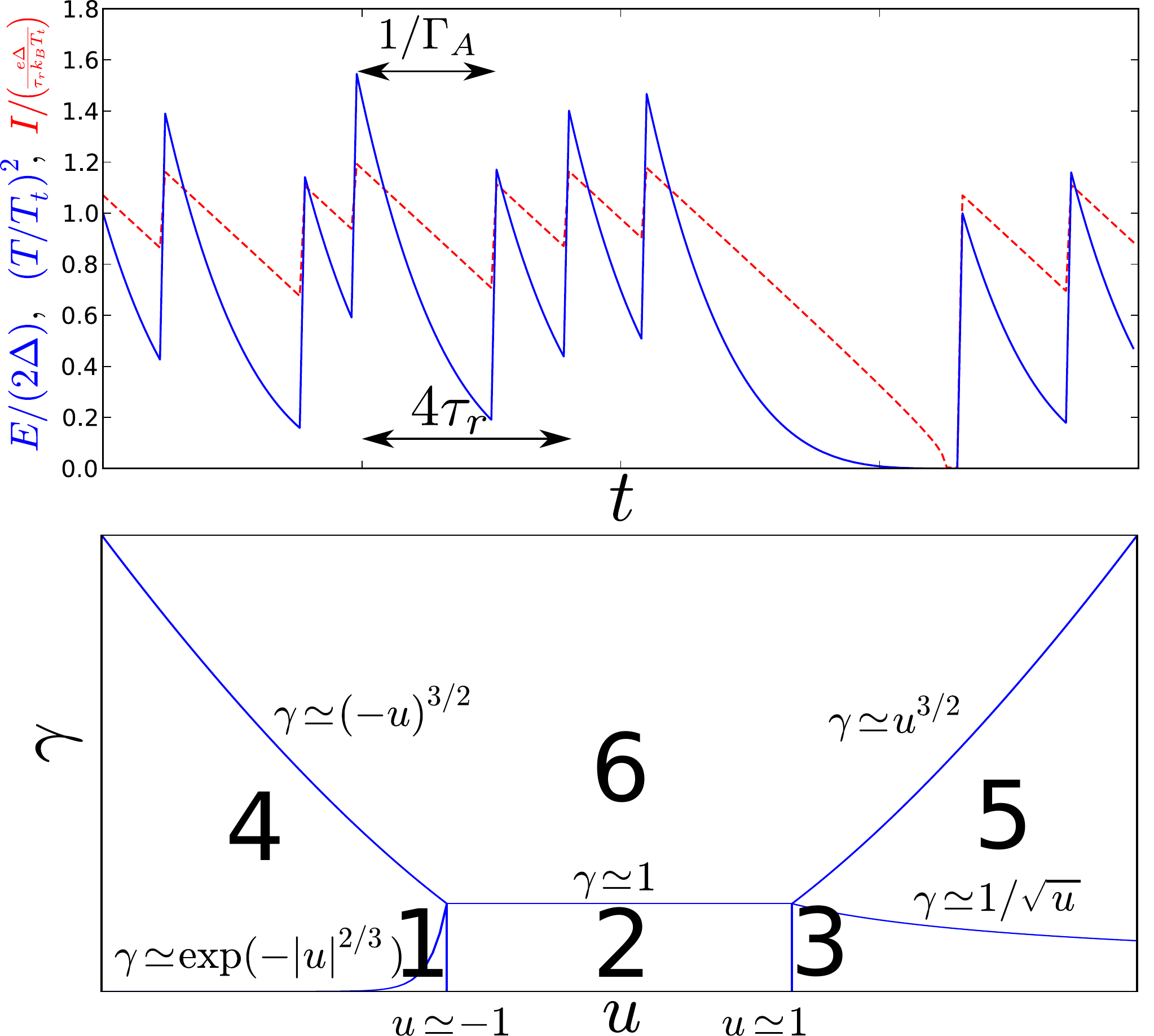}
	\caption{(color online) (top) Example timelines of energy (solid blue) and electric current (dashed red) for $eV=2\Delta$. Deterministic energy relaxation due to quasiparticle tunneling is characterized by a timescale $\tau_r$, relaxation time from $E=2\Delta$ to $E=0$ being $4\tau_r$ for $eV=2\Delta$. Andreev tunneling happens randomly at a rate $\Gamma_A$. Current and energy are related by Eq.~\eqref{eq:current}. (bottom) Voltage--Andreev rate "phase" diagram of the SINIS structure ($u=(eV/2-\Delta)/(k_BT_t)$, $\gamma=\Gamma_A\tau_r$). Regions 1, 2, and 3 exhibit non-Gaussian fluctuations.}
	\label{fig:timeline}
\end{figure}

Under these conditions, each Andreev tunneling substantially increases the temperature of the island. This increase is followed by the deterministic cooling at the timescale of $\tau_r$, and the temperature remains low until the next Andreev event (see an example timeline in Fig.~\ref{fig:timeline}). The distribution of temperature is mainly determined by the deterministic cooling and is strongly non-Gaussian. At an increased Andreev rate, $\Gamma_A\tau_r\gg1$, the deterministic cooling is too slow to substantially decrease the temperature between the Andreev events and the energy/temperature fluctuations become small and Gaussian.



With these assumptions, the time evolution of the probability distribution function for the total energy on the island, $\mathcal{P}(E)$, satisfies a Fokker--Plank equation
\begin{align}\label{eq:fp}
  \partial_t\mathcal{P}(E)=&-\partial_E\left(\dot{H}_\mr{qp}(E)\mathcal{P}(E)\right) \nonumber \\ &+\Gamma_A\left[\mathcal{P}(E-2\Delta)-\mathcal{P}(E)\right].
\end{align}
Here, the first term on the r.h.s.~describes the deterministic relaxation due to quasiparticles, $\dot{H}_\mr{qp}(E)$ being the energy flow of the quasiparticles, while the second term describes stochastic Andreev events with a rate $\Gamma_A$. We assume a symmetric setup \footnote{The effect of asymmetry is small, see \refcite{laakso11}.} and a bias voltage of $eV\approx2\Delta\gg k_BT$. Under these assumptions, the energy flow reads \cite{anghel01}
\begin{align}
  \dot{H}_\mr{qp}&=-\frac{\sqrt{2}g}{\hbar\pi}(k_BT)^{3/2}\Delta^{1/2}F((\Delta-eV/2)/(k_BT)),\!\! \\
  F(x)&=-\left[\Gamma\!\left(\frac{3}{2}\right)\!\mr{Li}_{3/2}(-e^{-x})+\Gamma\!\left(\frac{1}{2}\right)\!x\mr{Li}_{1/2}(-e^{-x})\right]. \nonumber 
\end{align}
$F(x)$ is positive at $x>0$, changing its sign at $x\approx-0.72$. Correspondingly, the quasiparticles always cool the island at $eV<2\Delta$. At $eV>2\Delta$, they fix the  temperature of the island to $k_B T\approx eV/2-\Delta$ \cite{giazotto06}. The Andreev rate is given by \cite{laakso11,belzig00}
\begin{equation}\label{eq:andreev}
 \Gamma_A=g\mathcal{F}\frac{\Delta}{4\pi\hbar}\ln\left[2\Delta/\max(\Delta-eV/2,k_BT)\right].
\end{equation}
It exhibits a weak logarithmic dependence on voltage and temperature that we disregard in the following.

Let us determine proper scales and corresponding dimensionless variables in the parameter region of interest. The natural scale for the total energy is $\Delta$, and we introduce a dimensionless energy, $\epsilon=E/(2\Delta)$. The island temperature in these units is given by $T=T_t\sqrt{\epsilon}$, and the natural scale for the bias voltage is correspondingly $u=(eV/2-\Delta)/(k_BT_t)$. In these units, the quasiparticle energy flow is
\begin{align}\label{eq:qp}
  \dot{H}_\mr{qp}&=-\frac{2\Delta}{\tau_r}\epsilon^{3/4}F(-u/\sqrt{\epsilon}), \\
  \frac{\hbar}{\tau_r}&\equiv\frac{g}{\pi}(k_BT_t)^{3/2}/(2\Delta)^{1/2}. \nonumber
\end{align}
Dimensionless time is naturally expressed as $\tau=t/\tau_r$. The condition for manifestly non-Gaussian fluctuations discussed, $\Gamma_A \tau_r \equiv \gamma \lesssim 1$, can thus be expressed in terms of the tunnel parameter $\mathcal{F}$, $\mathcal{F} \lesssim (k_BT_t/\Delta)^{3/2} \simeq (\delta/\Delta)^{3/4}$. It is precisely the existence of this additional small parameter of the order of $\delta/\Delta$ which allows us to reach the regime of non-Gaussian fluctuations.
 
The Fokker--Planck equation for the dimensionless variables reads
\begin{align}\label{eq:fp-dim}
 \partial_\tau\mathcal{P}(\epsilon)=&-\partial_\epsilon\left[-\epsilon^{3/4}F(-u/\sqrt{\epsilon})\mathcal{P}(\epsilon)\right] \nonumber \\ &+\gamma\left[\mathcal{P}(\epsilon-1)-\mathcal{P}(\epsilon)\right].
\end{align}
Its stationary solution depends on two dimensionless parameters $u$ and $\gamma$. There are six qualitatively different regimes, shown in Fig.~\ref{fig:timeline}, which we analyze briefly below.

If the stationary distribution is Gaussian, we can expand the difference $\mathcal{P}(\epsilon-1)-\mathcal{P}(\epsilon)$ in derivatives. In this case the average energy is determined from $v(\epsilon)\equiv \epsilon^{3/4}F(-u/\sqrt{\epsilon})=\gamma$, and the variance can be estimated by $\mr{Var}(\epsilon)\approx\left(\partial_\epsilon\ln v(\epsilon)\right)^{-1}$, $\mr{Var}(\epsilon)\gg1$ for a Gaussian distribution. There is a lower bound for the energy of the island where the cooling rate vanishes. It is given by the condition $v(\epsilon)=0$ with the result $\epsilon_\mr{min}=(1.39u)^2$ for $u>0$ and $\epsilon_\mr{min}=0$ for $u<0$.

Let us first concentrate on the case $|u|\lesssim1$, allowing us to approximate $F(-u/\sqrt{\epsilon})\approx F(0)\approx0.68$. In the Gaussian case we have $\langle\epsilon\rangle=\frac{3}{4}\mr{Var}(\epsilon)=(\gamma/F(0))^{4/3}$. In terms of temperature $\langle T\rangle\propto\gamma^{2/3}\Delta/k_B$, independent of $\delta$. This is regime 6 in Fig.~\ref{fig:timeline}. The crossover to non-Gaussian behavior, regime 2 in Fig.~\ref{fig:timeline}, happens when $\langle\epsilon\rangle\approx1$, i.e., $\gamma\approx F(0)$.

In the case $u\gg1$ the minimum energy is already larger than $\Delta$. However, at sufficiently small $\gamma$ the distribution near this minimum energy is non-Gaussian. We can approximate $v(\epsilon)$ by expanding it near $\epsilon_\mr{min}$. Introducing a new variable $\tilde{\epsilon}\equiv\epsilon-\epsilon_\mr{min}$ we obtain a simpler equation
\begin{equation}
 Cu^{-1/2}\partial_{\tilde{\epsilon}}\left[\tilde{\epsilon}\mathcal{P}(\tilde{\epsilon})\right]+\gamma\left[\mathcal{P}(\tilde{\epsilon}-1)-\mathcal{P}(\tilde{\epsilon})\right]=0,
\end{equation}
where $C\approx0.41$. In the Gaussian case (regime 5 in Fig.~\ref{fig:timeline}) we have $\langle\tilde{\epsilon}\rangle=\mr{Var}(\epsilon)=u^{1/2}\gamma/C$, with the crossover to non-Gaussian regime (regime 3 in Fig.~\ref{fig:timeline}) taking place at $\gamma\approx Cu^{-1/2}$, i.e., at a smaller value of $\gamma$ compared to the case $|u|\lesssim1$. Qualitatively new behavior takes place when $\langle\tilde{\epsilon}\rangle\approx\epsilon_\mr{min}$, that is, when $\gamma\gtrsim u^{3/2}$. In this case we can approximate $F(-u/\sqrt{\epsilon})\approx F(0)$, and recover the same behavior as in the case $|u|\lesssim1$ (regime 6).

Finally, in the case $|u|\gg1$, $u<0$, the quasiparticle rate is exponentially suppressed at low energies, $\sqrt{\epsilon}<|u|$. Let us assume that the distribution is concentrated near some value of energy $\epsilon_\ast\ll u^2$. Using the approximation $F(x)\approx\sqrt{\pi}xe^{-x}$ for $x\to\infty$, we have near $\epsilon\approx\epsilon_\mr\ast$
\begin{align}
 \sqrt{\pi}|u|\epsilon_\ast^{1/4}e^{-|u|/\sqrt{\epsilon_\ast}}\partial_{\tilde{\epsilon}}\left[e^{|u|\tilde{\epsilon}/(2\epsilon_\ast^{3/2})}\mathcal{P}(\tilde{\epsilon})\right]&\nonumber \\ +\gamma\left[\mathcal{P}(\tilde{\epsilon}-1)-\mathcal{P}(\tilde{\epsilon})\right]&=0,
\end{align}
where $\tilde{\epsilon}$ is the deviation from $\epsilon_\ast$. In the Gaussian case, we can estimate the variance as $\mr{Var}(\epsilon)\approx2\epsilon_\ast^{3/2}/|u|$. Comparing this with unity, we conclude that the Gaussian distribution is realized if $\epsilon_\ast>|u/2|^{2/3}\ll|u|$, i.e., if $\gamma\gtrsim|u|^{7/6}\exp(-2^{1/3}|u|^{2/3})$ (regime 4 in Fig.~\ref{fig:timeline}). For smaller $\gamma$ the distribution is non-Gaussian (regime 1 in Fig.~\ref{fig:timeline}). Upon further increase of $\gamma$, $\epsilon_\ast$ grows, allowing us to again approximate $F(-u/\sqrt{\epsilon})\approx F(0)$. It reaches $u^2$ at $\gamma\approx|u|^{3/2}$, taking us back to regime 6. This is similar to the situation with $u\gg1$.

\begin{figure}
	\includegraphics[width=0.85\columnwidth]{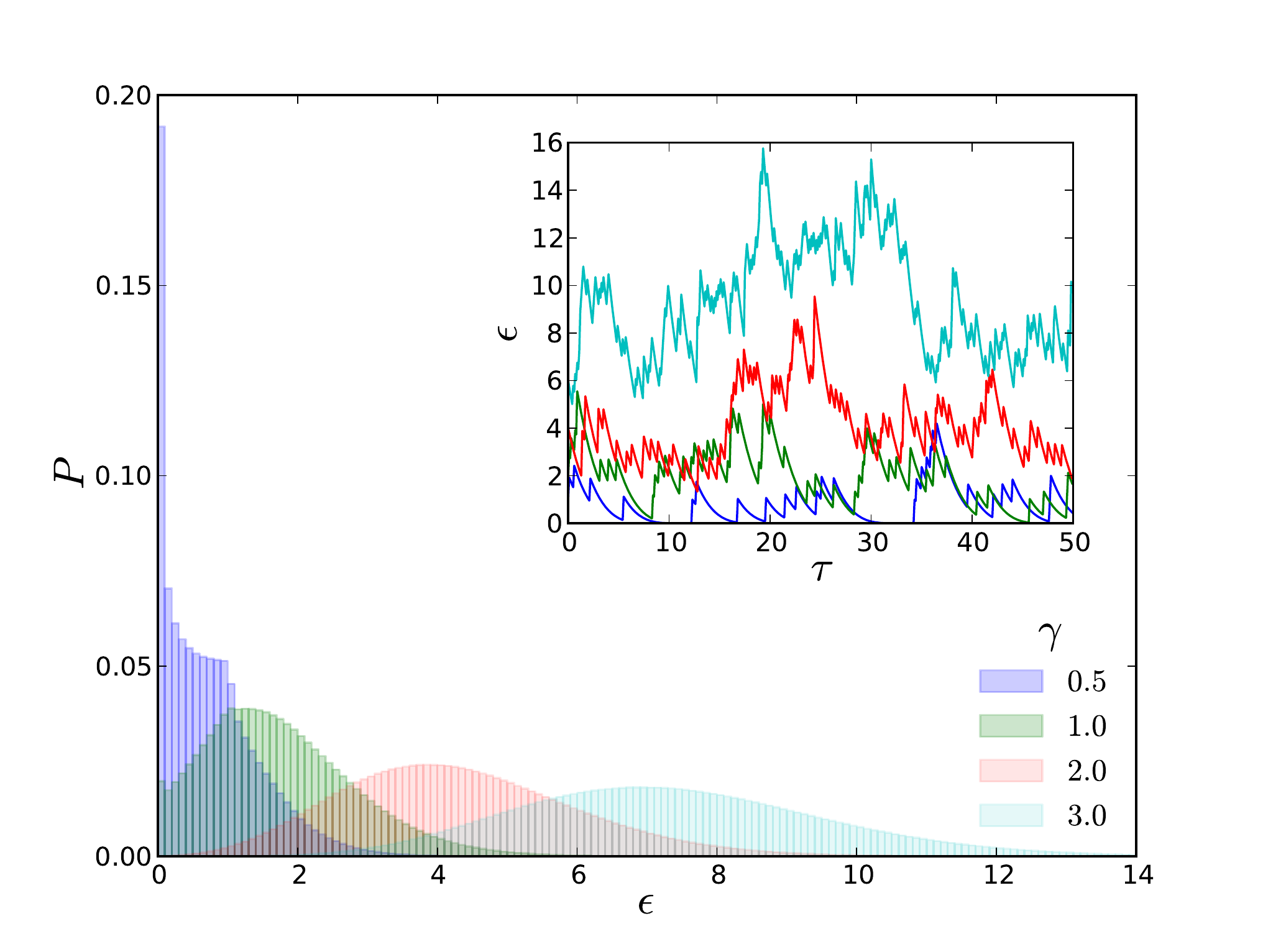}\vskip -0.2cm
	\includegraphics[width=0.85\columnwidth]{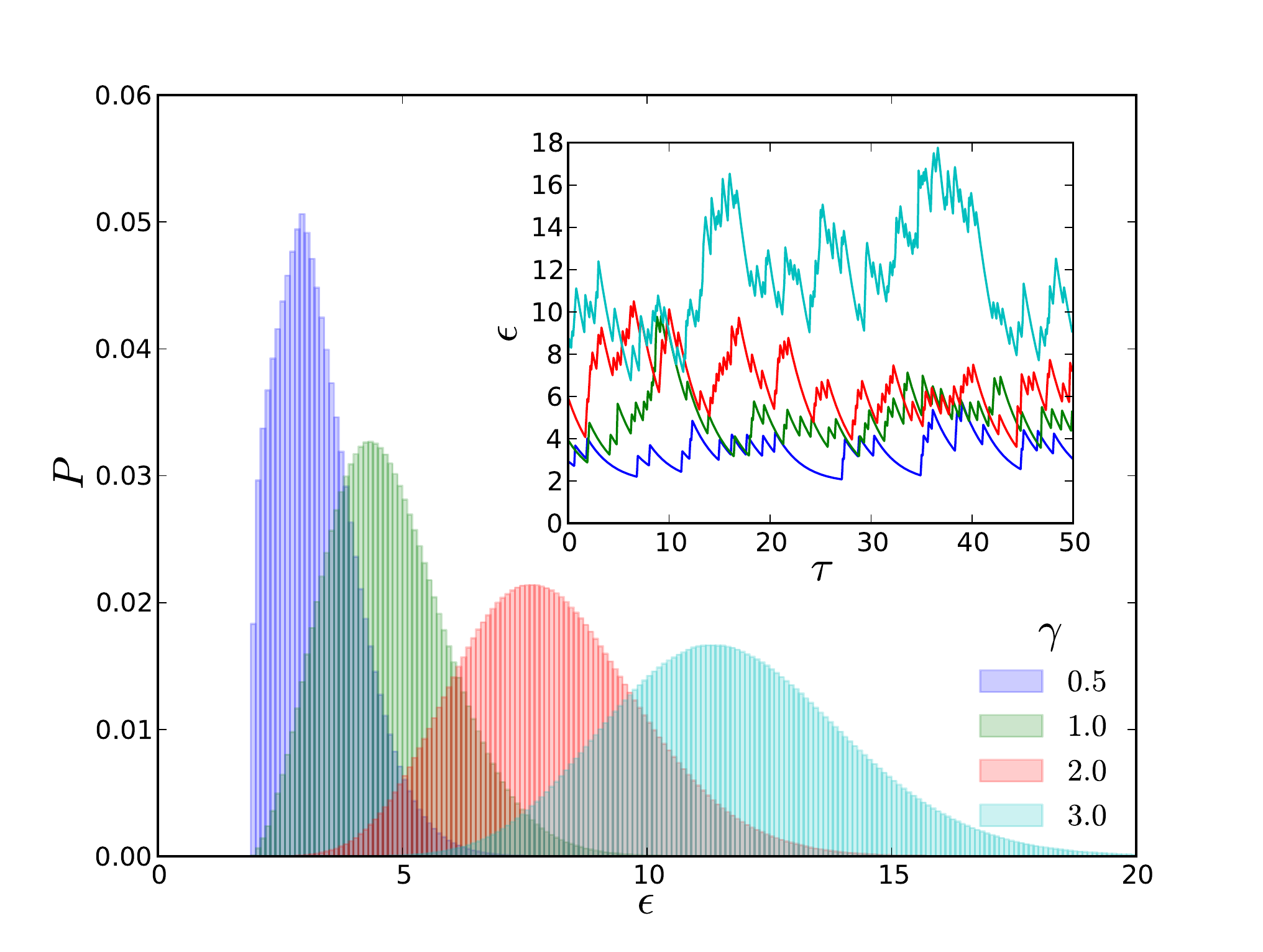}\vskip -0.2cm
	\includegraphics[width=0.85\columnwidth]{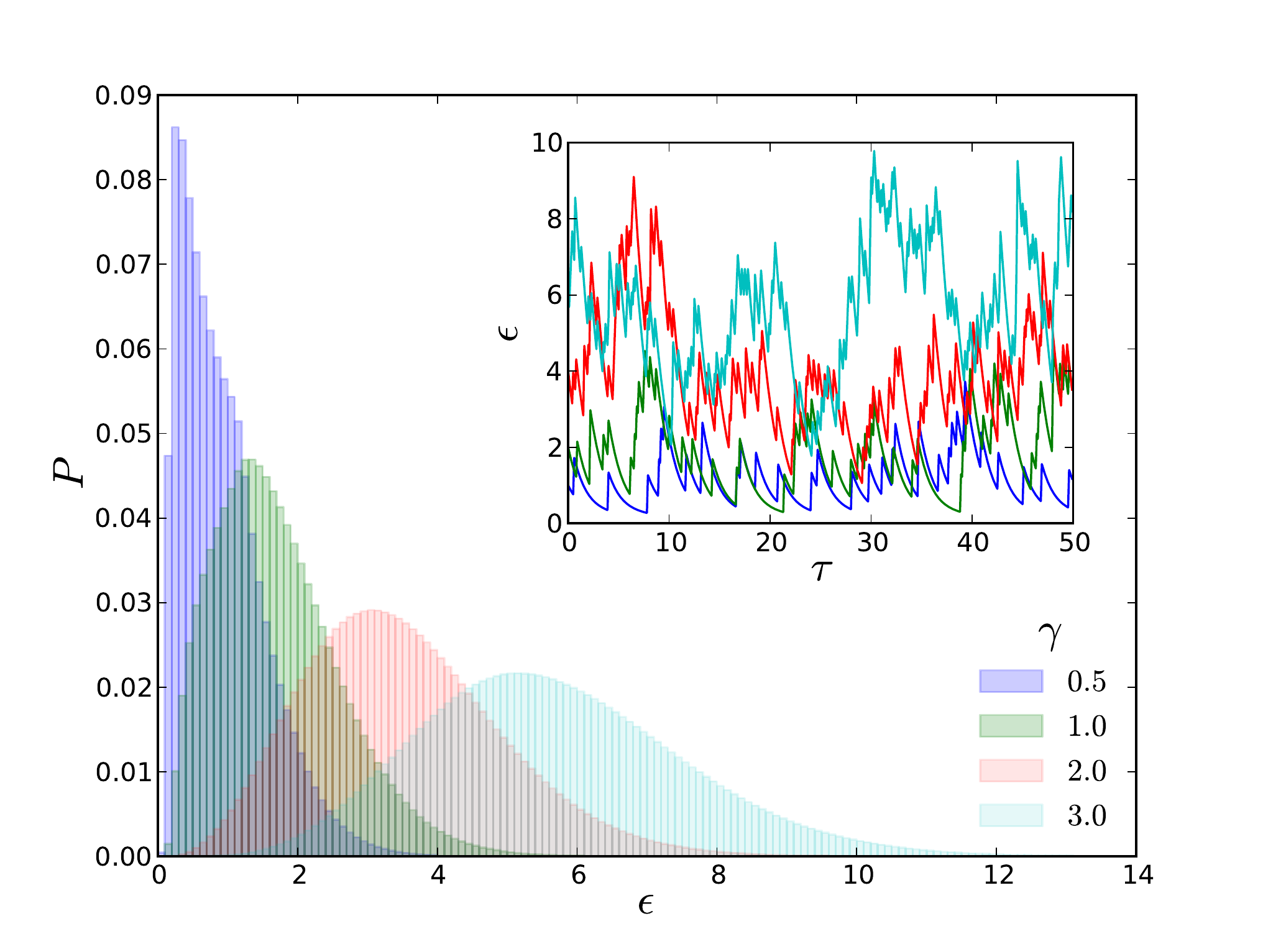}
	\caption{(color online) Probability distributions for the internal energy on the island for (top to bottom) $u=0$, $u=1$, and $u=-2$, and some values of $\gamma$. Insets show realizations of energy timelines for each value of $\gamma$. The bins of the histograms have a width of $0.1$.}
	\label{fig:edists}
\end{figure}
The probability distributions for energy, calculated numerically from Eq.~\eqref{eq:fp-dim}, are shown in Fig.~\ref{fig:edists}. We choose three different values of $u$ corresponding to the three non-Gaussian regimes, and vary $\gamma$ to observe the crossover to Gaussian behavior.

The topmost figure shows the probabilities for $u=0$. The distributions for the two smallest values of $\gamma$ are strongly non-Gaussian, belonging to regime 2 in Fig.~\ref{fig:timeline}. For larger $\gamma$ the average energy grows, being proportional to $\gamma^{4/3}$. The inset in the figure shows some realizations of energy timelines corresponding to the distributions. For small $\gamma$ one can clearly see the difference between Andreev events, which are Poisson distributed and increase the energy by $2\Delta$, and quasiparticle processes, which steadily cool down the island. For large $\gamma$ the energy-increasing and energy-decreasing processes look the same, characteristic of a Gaussian distribution. As seen from the timeline, the energy on the island can reach zero and stay there for a finite time, implying that the temperature goes to absolute zero. In reality, the temperature would be about $\delta/k_B\ll T_t$ when only a few particles are excited on the island, a situation outside the scope of the present model.


The middle figure corresponds to $u=1$. Since $u>0$, the minimum energy is finite, $\epsilon_\mr{min}\approx1.93$. The distribution for the smallest value of $\gamma$ is clearly non-Gaussian, belonging to regime 3 in Fig.~\ref{fig:timeline}. For larger $\gamma$ the average energy grows linearly.

The last figure shows the probabilities for $u=-2$. Again, the distribution for the smallest value of $\gamma$ is non-Gaussian, belonging to regime 1 in Fig.~\ref{fig:timeline}. Upon an increase of $\gamma$ the main body of the distribution transforms into a Gaussian one and the average energy grows. The tails of the distribution deviate from Gaussian even for the largest $\gamma$ shown.

Since the electric current through the SINIS structure is sensitive to the instantaneous temperature on the island, the statistics of internal energy and temperature can be gathered by monitoring the instantaneous electric current. The necessary relation is given by
\begin{equation}\label{eq:current}
 I=-\frac{e}{\tau_r}\frac{\Delta}{k_BT_t}\epsilon^{1/4}\Gamma\left(\frac{1}{2}\right)\mr{Li}_{1/2}(-e^{u/\sqrt{\epsilon}}),
\end{equation}
which can be numerically inverted. The time scale of the fluctuations, $\tau_r$, is of the order of $10\;\mr{ns}$, and the corresponding current scale, $e\Delta/(\tau_rk_BT_t)$, of the order of $1\;\mr{nA}$ for $\delta=10^{-4}\;\mr{K}\times k_B$, $\Delta=1\;\mr{K}\times k_B$, $T_t=10\;\mr{mK}$ and $g=3$, corresponding to a copper island of volume $\mathcal{V}=0.01\;(\mu\mr{m})^3$ connected to aluminum leads.


In summary, we suggest a prototypical setup utilizing a voltage biased SINIS tunnel structure for the detection of energy and temperature fluctuations out of equilibrium. We have identified parameter regimes which exhibit fluctuations that are both strong and non-Gaussian, and we suggest that the measurement of instantaneous electric current could be used to gather the statistics of these fluctuating quantities. We have also described a possibility to realize Maxwell--demon-like feedback schemes.

\acknowledgments
We thank J.~P.~Pekola for useful discussions. This work was supported by the Finnish Academy of Science and Letters, the Academy of Finland, and the European Research Council (Grant No.~240362-Heattronics).


\begin{thebibliography}{10}%
\makeatletter
\providecommand \@ifxundefined [1]{%
 \ifx #1\undefined \expandafter \@firstoftwo
 \else \expandafter \@secondoftwo
\fi
}%
\providecommand \@ifnum [1]{%
 \ifnum #1\expandafter \@firstoftwo
 \else \expandafter \@secondoftwo
\fi
}%
\providecommand \enquote [1]{``#1''}%
\providecommand \bibnamefont  [1]{#1}%
\providecommand \bibfnamefont [1]{#1}%
\providecommand \citenamefont [1]{#1}%
\providecommand\href[0]{\@sanitize\@href}%
\providecommand\@href[1]{\endgroup\@@startlink{#1}\endgroup\@@href}%
\providecommand\@@href[1]{#1\@@endlink}%
\providecommand \@sanitize [0]{\begingroup\catcode`\&12\catcode`\#12\relax}%
\@ifxundefined \pdfoutput {\@firstoftwo}{%
 \@ifnum{\z@=\pdfoutput}{\@firstoftwo}{\@secondoftwo}%
}{%
 \providecommand\@@startlink[1]{\leavevmode}%
 \providecommand\@@endlink[0]{}%
}{%
 \providecommand\@@startlink[1]{%
  \leavevmode
  \pdfstartlink
   attr{/Border[0 0 1 ]/H/I/C[0 1 1]}%
   user{/Subtype/Link/A<</Type/Action/S/URI/URI(#1)>>}%
  \relax
 }%
 \providecommand\@@endlink[0]{\pdfendlink}%
}%
\providecommand \url  [0]{\begingroup\@sanitize \@url }%
\providecommand \@url [1]{\endgroup\@href {#1}{\urlprefix}}%
\providecommand \urlprefix [0]{URL }%
\providecommand \Eprint[0]{\href }%
\@ifxundefined \urlstyle {%
  \providecommand \doi [1]{doi:\discretionary{}{}{}#1}%
}{%
  \providecommand \doi [0]{doi:\discretionary{}{}{}\begingroup
  \urlstyle{rm}\Url }%
}%
\providecommand \doibase [0]{http://dx.doi.org/}%
\providecommand \Doi[1]{\href{\doibase#1}}%
\providecommand \bibAnnote [3]{%
  \BibitemShut{#1}%
  \begin{quotation}\noindent
    \textsc{Key:}\ #2\\\textsc{Annotation:}\ #3%
  \end{quotation}%
}%
\providecommand \bibAnnoteFile [2]{%
  \IfFileExists{#2}{\bibAnnote {#1} {#2} {\input{#2}}}{}%
}%
\providecommand \typeout [0]{\immediate \write \m@ne }%
\providecommand \selectlanguage [0]{\@gobble}%
\providecommand \bibinfo [0]{\@secondoftwo}%
\providecommand \bibfield [0]{\@secondoftwo}%
\providecommand \translation [1]{[#1]}%
\providecommand \BibitemOpen[0]{}%
\providecommand \bibitemStop [0]{}%
\providecommand \bibitemNoStop [0]{.\EOS\space}%
\providecommand \EOS [0]{\spacefactor3000\relax}%
\providecommand \BibitemShut [1]{\csname bibitem#1\endcsname}%
\bibitem{reichl98}%
  \BibitemOpen
  \bibfield{author}{%
  \bibinfo {author} {\bibfnamefont{L.~E.}\ \bibnamefont{Reichl}},\ }%
  \emph{\bibinfo {title} {{A Modern Course in Statistical Physics}}},\ \bibinfo
  {edition} {2nd}\ ed.\ (\bibinfo {publisher} {John Wiley \& Sons, Inc.},\
  \bibinfo {year} {1998})%
  \bibAnnoteFile{NoStop}{reichl98}%
\bibitem{heikkila09}%
  \BibitemOpen
  \bibfield{author}{%
  \bibinfo {author} {\bibfnamefont{T.~T.}\ \bibnamefont{Heikkil{\"a}}}\ and\
  \bibinfo {author} {\bibfnamefont{Y.~V.}\ \bibnamefont{Nazarov}},\ }%
  \bibfield{journal}{%
  \Doi{10.1103/PhysRevLett.102.130605}{\bibinfo {journal} {Phys. Rev. Lett.}}\
  }%
  \textbf{\bibinfo {volume} {102}},\ \bibinfo {pages} {130605} (\bibinfo {year}
  {2009})%
  \bibAnnoteFile{NoStop}{heikkila09}%
\bibitem{laakso10}%
  \BibitemOpen
  \bibfield{author}{%
  \bibinfo {author} {\bibfnamefont{M.~A.}\ \bibnamefont{Laakso}}, \bibinfo
  {author} {\bibfnamefont{T.~T.}\ \bibnamefont{Heikkil{\"a}}},\ and\ \bibinfo
  {author} {\bibfnamefont{Y.~V.}\ \bibnamefont{Nazarov}},\ }%
  \bibfield{journal}{%
  \Doi{10.1103/PhysRevLett.104.196805}{\bibinfo {journal} {Phys. Rev. Lett.}}\
  }%
  \textbf{\bibinfo {volume} {104}},\ \bibinfo {pages} {196805} (\bibinfo {year}
  {2010})%
  \bibAnnoteFile{NoStop}{laakso10}%
\bibitem{laakso10b}%
  \BibitemOpen
  \bibfield{author}{%
  \bibinfo {author} {\bibfnamefont{M.~A.}\ \bibnamefont{Laakso}}, \bibinfo
  {author} {\bibfnamefont{T.~T.}\ \bibnamefont{Heikkil{\"a}}},\ and\ \bibinfo
  {author} {\bibfnamefont{Y.~V.}\ \bibnamefont{Nazarov}},\ }%
  \bibfield{journal}{%
  \Doi{10.1103/PhysRevB.82.205316}{\bibinfo {journal} {Phys. Rev. B}}\ }%
  \textbf{\bibinfo {volume} {82}},\ \bibinfo {pages} {205316} (\bibinfo {year}
  {2010})%
  \bibAnnoteFile{NoStop}{laakso10b}%
\bibitem{serreli07}%
  \BibitemOpen
  \bibfield{author}{%
  \bibinfo {author} {\bibfnamefont{V.}~\bibnamefont{Serreli}}, \bibinfo
  {author} {\bibfnamefont{C.-F.}\ \bibnamefont{Lee}}, \bibinfo {author}
  {\bibfnamefont{E.~R.}\ \bibnamefont{Kay}},\ and\ \bibinfo {author}
  {\bibfnamefont{D.~A.}\ \bibnamefont{Leigh}},\ }%
  \bibfield{journal}{%
  \Doi{10.1038/nature05452}{\bibinfo {journal} {Nature}}\ }%
  \textbf{\bibinfo {volume} {445}},\ \bibinfo {pages} {523} (\bibinfo {year}
  {2007})%
  \bibAnnoteFile{NoStop}{serreli07}%
\bibitem{horowitz11}%
  \BibitemOpen
  \bibfield{author}{%
  \bibinfo {author} {\bibfnamefont{M.}~\bibnamefont{Horowitz}}\ and\ \bibinfo
  {author} {\bibfnamefont{J.~M.~R.}\ \bibnamefont{Parrondo}},\ }%
  \bibfield{journal}{%
  \Doi{10.1209/0295-5075}{\bibinfo {journal} {Europhys. Lett.}}\ }%
  \textbf{\bibinfo {volume} {95}},\ \bibinfo {pages} {10005} (\bibinfo {year}
  {2011})%
  \bibAnnoteFile{NoStop}{horowitz11}%
\bibitem{pekola07}%
  \BibitemOpen
  \bibfield{author}{%
  \bibinfo {author} {\bibfnamefont{J.}~\bibnamefont{Pekola}}\ and\ \bibinfo
  {author} {\bibfnamefont{F.}~\bibnamefont{Hekking}},\ }%
  \bibfield{journal}{%
  \Doi{10.1103/PhysRevLett.98.210604}{\bibinfo {journal} {Phys. Rev. Lett.}}\
  }%
  \textbf{\bibinfo {volume} {98}},\ \bibinfo {pages} {210604} (\bibinfo {year}
  {2007})%
  \bibAnnoteFile{NoStop}{pekola07}%
\bibitem{Note1}%
  \BibitemOpen
  \bibinfo {note} {For the duration of the electron--phonon scattering time,
  assumed long in this paper.}%
  \bibAnnoteFile{Stop}{Note1}%
\bibitem{nahum94}%
  \BibitemOpen
  \bibfield{author}{%
  \bibinfo {author} {\bibfnamefont{M.}~\bibnamefont{Nahum}}, \bibinfo {author}
  {\bibfnamefont{T.~M.}\ \bibnamefont{Eiles}},\ and\ \bibinfo {author}
  {\bibfnamefont{J.~M.}\ \bibnamefont{Martinis}},\ }%
  \bibfield{journal}{%
  \Doi{10.1063/1.112456}{\bibinfo {journal} {Appl. Phys. Lett.}}\ }%
  \textbf{\bibinfo {volume} {65}},\ \bibinfo {pages} {3123} (\bibinfo {year}
  {1994})%
  \bibAnnoteFile{NoStop}{nahum94}%
\bibitem{leivo96}%
  \BibitemOpen
  \bibfield{author}{%
  \bibinfo {author} {\bibfnamefont{M.~M.}\ \bibnamefont{Leivo}}, \bibinfo
  {author} {\bibfnamefont{J.~P.}\ \bibnamefont{Pekola}},\ and\ \bibinfo
  {author} {\bibfnamefont{D.~V.}\ \bibnamefont{Averin}},\ }%
  \bibfield{journal}{%
  \Doi{10.1063/1.115651}{\bibinfo {journal} {Appl. Phys. Lett.}}\ }%
  \textbf{\bibinfo {volume} {68}},\ \bibinfo {pages} {1996} (\bibinfo {year}
  {1996})%
  \bibAnnoteFile{NoStop}{leivo96}%
\bibitem{giazotto06}%
  \BibitemOpen
  \bibfield{author}{%
  \bibinfo {author} {\bibfnamefont{F.}~\bibnamefont{Giazotto}} \textit{et al.},\ }%
  \bibfield{journal}{%
  \Doi{10.1103/RevModPhys.78.217}{\bibinfo {journal} {Rev. Mod. Phys.}}\ }%
  \textbf{\bibinfo {volume} {78}},\ \bibinfo {pages} {217} (\bibinfo {year}
  {2006})%
  \bibAnnoteFile{NoStop}{giazotto06}%
\bibitem{hekking93}%
  \BibitemOpen
  \bibfield{author}{%
  \bibinfo {author} {\bibfnamefont{F.~W.~J.}\ \bibnamefont{Hekking}}\ and\
  \bibinfo {author} {\bibfnamefont{Y.~V.}\ \bibnamefont{Nazarov}},\ }%
  \bibfield{journal}{%
  \Doi{10.1103/PhysRevLett.71.1625}{\bibinfo {journal} {Phys. Rev. Lett.}}\ }%
  \textbf{\bibinfo {volume} {71}},\ \bibinfo {pages} {1625} (\bibinfo {year}
  {1993})%
  \bibAnnoteFile{NoStop}{hekking93}%
\bibitem{rajauria08}%
  \BibitemOpen
  \bibfield{author}{%
  \bibinfo {author} {\bibfnamefont{S.}~\bibnamefont{Rajauria}} \textit{et al.},\ }%
  \bibfield{journal}{%
  \Doi{10.1103/PhysRevLett.100.207002}{\bibinfo {journal} {Phys. Rev. Lett.}}\
  }%
  \textbf{\bibinfo {volume} {100}},\ \bibinfo {pages} {207002} (\bibinfo {year}
  {2008})%
  \bibAnnoteFile{NoStop}{rajauria08}%
\bibitem{belzig00}%
  \BibitemOpen
  \bibfield{author}{%
  \bibinfo {author} {\bibfnamefont{W.}~\bibnamefont{Belzig}}, \bibinfo {author}
  {\bibfnamefont{A.}~\bibnamefont{Brataas}}, \bibinfo {author}
  {\bibfnamefont{Y.~V.}\ \bibnamefont{Nazarov}},\ and\ \bibinfo {author}
  {\bibfnamefont{G.~E.~W.}\ \bibnamefont{Bauer}},\ }%
  \bibfield{journal}{%
  \Doi{10.1103/PhysRevB.62.9726}{\bibinfo {journal} {Phys. Rev. B}}\ }%
  \textbf{\bibinfo {volume} {62}},\ \bibinfo {pages} {9726} (\bibinfo {year}
  {2000})%
  \bibAnnoteFile{NoStop}{belzig00}%
\bibitem{maisi11}%
  \BibitemOpen
  \bibfield{author}{%
  \bibinfo {author} {\bibfnamefont{V.~F.}\ \bibnamefont{Maisi}} \textit{et al.},\ }%
  \bibfield{journal}{%
  \Doi{10.1103/PhysRevLett.106.217003}{\bibinfo {journal} {Phys. Rev. Lett.}}\
  }%
  \textbf{\bibinfo {volume} {106}},\ \bibinfo {pages} {217003} (\bibinfo {year}
  {2011})%
  \bibAnnoteFile{NoStop}{maisi11}%
\bibitem{greibe11}%
  \BibitemOpen
  \bibfield{author}{%
  \bibinfo {author} {\bibfnamefont{T.}~\bibnamefont{Greibe}} \textit{et al.},\ }%
  \bibfield{journal}{%
  \Doi{10.1103/PhysRevLett.106.097001}{\bibinfo {journal} {Phys. Rev. Lett.}}\
  }%
  \textbf{\bibinfo {volume} {106}},\ \bibinfo {pages} {97001} (\bibinfo {year}
  {2011})%
  \bibAnnoteFile{NoStop}{greibe11}%
\bibitem{Note2}%
  \BibitemOpen
  \bibinfo {note} {This is not a very limiting requirement, since the backflow
  of heat from the reservoirs at temperature $T_S$ is exponentially small for
  $k_BT_S\ll \Delta $, $\propto \protect \sqrt {T_S}\protect \qopname \relax
  o{exp}[-\Delta /(k_BT_S)]$ \cite {anghel01}.}%
  \bibAnnoteFile{Stop}{Note2}%
\bibitem{Note3}%
  \BibitemOpen
  \bibinfo {note} {The effect of asymmetry is small, see Ref.~\protect
  \rev@citealp {laakso11}.}%
  \bibAnnoteFile{Stop}{Note3}%
\bibitem{anghel01}%
  \BibitemOpen
  \bibfield{author}{%
  \bibinfo {author} {\bibfnamefont{D.~V.}\ \bibnamefont{Anghel}}\ and\ \bibinfo
  {author} {\bibfnamefont{J.~P.}\ \bibnamefont{Pekola}},\ }%
  \bibfield{journal}{%
  \Doi{10.1023/A:1017589828739}{\bibinfo {journal} {J. Low Temp. Phys}}\ }%
  \textbf{\bibinfo {volume} {123}},\ \bibinfo {pages} {197} (\bibinfo {year}
  {2001})%
  \bibAnnoteFile{NoStop}{anghel01}%
\bibitem{laakso11}%
  \BibitemOpen
  \bibfield{author}{%
  \bibinfo {author} {\bibfnamefont{M.~A.}\ \bibnamefont{Laakso}}, \bibinfo
  {author} {\bibfnamefont{T.~T.}\ \bibnamefont{Heikkil{\"a}}},\ and\ \bibinfo
  {author} {\bibfnamefont{Y.~V.}\ \bibnamefont{Nazarov}},\ }%
  (\bibinfo {year}
  {2011}),\ \Eprint{http://arxiv.org/abs/arXiv:1110.6726v1}{arXiv:1110.6726v1}%
  \bibAnnoteFile{NoStop}{laakso11}%
\end{thebibliography}

%

\end{document}